\documentclass[a4paper,superscriptaddress,aps,prl,twocolumn,floatfix,citeautoscript]{revtex4}
\usepackage{graphicx}
\usepackage{amsmath}
\usepackage{natbib}
\usepackage{mciteplus}
\usepackage{color}
\usepackage{float}
\usepackage[normalem]{ulem}

\newcommand{\vk}{\mathbf{k}}

\begin{document}
\title{Accurate optical spectra of solids from pure time-dependent density-functional theory}

\newcommand{\lpt}{Laboratoire de Physique Th\'eorique, Universit\'e de Toulouse, CNRS, UPS and and European Theoretical Spectroscopy Facility, France}
\newcommand{\lcpq}{Laboratoire de Chimie et Physique Quantiques, Universit\'e de Toulouse, CNRS, UPS and and European Theoretical Spectroscopy Facility, France}


\author{Sarah Cavo}
\affiliation{\lpt}

\author{J. A. Berger}
\affiliation{\lcpq}

\author{Pina Romaniello}
\affiliation{\lpt} 

\date{\today}


\begin{abstract}

We present accurate optical spectra of semiconductors and insulators within a pure Kohn-Sham time-dependent density-functional approach.
In particular, we show that the onset of the absorption is well reproduced when comparing to experiment.
No empirical information nor a theory beyond Kohn-Sham density-functional theory, such as $GW$, is invoked to correct the Kohn-Sham gap. Our approach relies on the link between the exchange-correlation kernel of time-dependent density functional theory and the derivative discontinuity of ground-state density-functional theory. We show explicitly how to relate these two quantities.
We illustrate the accuracy and simplicity of our approach by applying it to various semiconductors (Si, GaP, GaAs) and wide-gap insulators (C, LiF, Ar).
%


\end{abstract}
\maketitle
%
Time-dependent-density functional theory (TDDFT)~\cite{RungeGross} has become, over the years, one of the few well-established first-principles' approaches to describe time-dependent phenomena for a large variety of systems, both in the linear-response regime and beyond (see, e.g., Refs~[\onlinecite{Ullrich_book2013, Maitra_JCP2016}] and references therein).
In the last two decades TDDFT has been increasingly applied to solids, and in particular to the calculation of the optical absorption spectra. Optical experiments in general are very useful tools to investigate and characterize condensed-matter systems; it is hence desirable to develop efficient and reliable theoretical approaches to complement experiment.

Within TDDFT the description of optical spectra depends crucially on the exchange-correlation (xc) kernel $f_{xc}$ which relates the response of the Kohn-Sham system to a small perturbation to the response of the true system.
Thanks to the numerical efficiency of TDDFT, it is desirable that a simple but accurate xc kernel is available for the calculation of optical spectra.
It is well-known that traditional xc kernels, such as the random-phase approximation (RPA), i.e., $f_{xc}=0$, and the adiabatic local-density approximation (ALDA)~\cite{ZangwillSoven}, fail to describe two important features of optical spectra: 
1) excitonic effects and 2) the absorption onset.
While excitonic effects can nowadays be described accurately for various systems with a several xc kernels ~\cite{bootstrap,Trevisanutto,santiago,Berger_2015,Yang_2015,Terentjev_2018}, 
the correct description of the absorption onset within TDDFT remains an unsolved problem.

The starting point for a TDDFT calculation is the Kohn-Sham band structure.
As is well known, the Kohn-Sham band gap is, in general, estimated to be much smaller than the fundamental gap, i.e., the difference between the ionization potential and the electron affinity.~\cite{Godby_1986,Godby_1987,Gruning_2006}.
Therefore, since TDDFT should give the exact absorption spectra, the TDDFT xc kernel has the difficult task to ensure that there is no absorption below the optical gap even though transitions between Kohn-Sham valence and conduction bands are available.
Indeed, optical spectra obtained with all currently available xc kernels show absorption at energies close to the Kohn-Sham band gap, thereby severely underestimating the absorption onset.

The standard approach to circumvent this problem is to add a scissors operator~\cite{Levine_1989} to the Kohn-Sham Hamiltonian. The shift parameter 
is either obtained from experiment or from a method that goes beyond KS-DFT, such as $GW$~\cite{HedinPR,Aryasetiawan_1998,AulburJonssonWilkins,Reining_2017, Golze_2019}, or generalised KS-DFT with a hybrid functional, such as those based on a screened Coulomb interaction 
\cite{HSE03,HSE06,Matsushita_PRB2011,Galli_PRB2016}.
The former approach is unsatisfactory because it is empirical while the latter approach is unsatisfactory both conceptually and numerically because i) one has to go to a theory outside of the KS-DFT framework, i.e., $GW$ (or a hybrid functional), in order to obtain spectra that are comparable with those observed in experiment, and ii) a $GW$ (or a hybrid functional) calculation is much more expensive than a pure KS-DFT calculation. Eliminating the intermediate $GW$ step yields a fully coherent theory. Moreover, it leads to a large speed up of calculations, making the optical spectra of larger systems accessible. We note that one could use an \textit{ad hoc} meta-GGA functional specifically constructed to obtain a KS gap similar to the fundamental gap. However, this is not in agreement with fundamental DFT theory because the KS gap is not equal to the fundamental gap, the difference between the two being the derivative discontinuity.



The TDDFT xc kernel $f_{xc}$ can be written exactly as \cite{Sottile_PRL2003}
\begin{align}
f_{xc}(1,2) &= \underbrace{\chi^{-1}_{\text{KS}}(1,2) - \chi_0^{-1}(1,2)}_{f_{xc}^{(1)}}
\notag \\ &
\underbrace{-i\int d345 \chi_0^{-1} (1,5)G(5,3) G(4,5) \frac{\delta\Sigma(3,4)}{\delta\rho(2)}}_{f_{xc}^{(2)}}
\label{Eqn:fxc_exact}
\end{align}
where $\chi_{\text{KS}}$ and $\chi_0=-iGG$ are the Kohn-Sham and independent quasiparticle polarizability, respectively, and $G(1,2)$ and $\Sigma(1,2)$ are 
the one-body Green function and the self-energy, respectively. 
The collective index $(1)=(\mathbf{x},t)=(\mathbf{r},s,t)$ contains the space, spin and time coordinates.
The xc kernel written in Eq.~\eqref{Eqn:fxc_exact} clearly exhibits two distinct parts.
The first part $f_{xc}^{(1)} = \chi^{-1}_{\text{KS}} - \chi_0^{-1}$ only involves
independent (quasi-)particles, and, therefore,
is responsible for the shift of the Kohn-Sham band gap to the fundamental gap,
while the second part, $f_{xc}^{(2)}$, which includes the electron-hole interaction, accounts for the excitonic effects.
In general, both terms are required to guarantee a correct onset of the absorption, unless the exciton binding energy is small, in which case $f_{xc}^{(1)}$ is sufficient.
Although Eq.~\eqref{Eqn:fxc_exact} clearly distinguishes these two parts it is not useful in practical applications since it would require the calculation of $G$.
From the above discussion one would expect a link between $f_{xc}^{(1)}$ and the derivative discontinuity of ground-state DFT~\cite{Perdew_1982,Perdew_1983}, which is defined as the difference between the fundamental gap and the Kohn-Sham gap. One of the goals of this work is to make this link explicit.

In order to obtain this formal link we first generalize the two-point KS polarizability to four points:~\cite{OnidaReiningRubio}
\begin{align}
{^4}\chi_{\text{KS}}(\mathbf{x}_1,\mathbf{x}_2,\mathbf{x}_3,\mathbf{x}_4,\omega) & = \sum_{i,j}(f_j - f_i) 
\notag \\ & \times
\frac{\phi_i(\mathbf{x}_1)\phi_j(\mathbf{x}_2)\phi^*_j(\mathbf{x}_3)\phi^*_i(\mathbf{x}_4)}{\omega - (\epsilon_i -\epsilon_j) +i\eta},
\end{align}
where $\phi_i$ is a KS spinorbital, $\epsilon_i$ its energy, $f_i$ its occupation (0 and 1 for unoccupied and occupied orbitals, respectively), and $\eta$ is a positive infinitesimal that ensures causality. For the solids we study here $i$ and $a$ are multi-indices composed of a band index (comprising the spin) and a Bloch vector, $\vk$ and $\vk'$, respectively. We note that, although the final goal is the description of optical absorption for which $\vk' \to \vk$, the discussion below is completely general, i.e., $\vk \neq \vk'$.
The usual two-point KS polarizability is retrieved from the diagonal part of $\chi^{(4)}_{\text{KS}}$ as
\begin{equation}
\chi_{\text{KS}}(\mathbf{x}_1,\mathbf{x}_2,\omega) = {^4}\chi_{\text{KS}}(\mathbf{x}_1,\mathbf{x}_2,\mathbf{x}_1,\mathbf{x}_2,\omega).
\label{Eqn:contraction}
\end{equation}
%

We can express $\chi^{(4)}_{\text{KS}}$ in the Kohn-Sham basis by using the following basis transformation
\begin{align}
^4\chi^{\quad[n_4n_2]}_{KS[n_1n_3]}(\omega)&=\int d\mathbf{x}_1\mathbf{x}_2\mathbf{x}_3\mathbf{x}_4 {^4}\chi_{\text{KS}}(\mathbf{x}_1,\mathbf{x}_2,\mathbf{x}_3,\mathbf{x}_4,\omega)
\notag
\\ & \times \phi^*_{n_1}(\mathbf{x}_1)\phi^*_{n_2}(\mathbf{x}_2)\phi_{n_3}(\mathbf{x}_3)\phi_{n_4}(\mathbf{x}_4).
\label{Eqn:basis_trans}
\end{align}
This yields a $2M\times 2M$ diagonal matrix with $M$ the number of KS excitations.
It is schematically given by~\cite{Rohlfing_2000}
\begin{equation}
^4\chi_{\text{KS}}(\omega) \!=\!
\left(
\begin{array}{cccccc}
\!\frac{1}{\omega-\omega_1}\! &  &  &  & & \\
&  \ddots  & & & & \\
&  & \!\frac{1}{\omega-\omega_M}\! & & & \\
& &  & \!-\!\frac{1}{\omega+\omega_1} \!& &  \\
&  & & & \ddots &\\
&  & & & & \!-\!\frac{1}{\omega+\omega_M}\!
\end{array}
\right)
\end{equation}
%
%
where $\omega_i$ is a KS excitation energy, i.e., a pole of $\chi_{\text{KS}}(\omega)$, and the matrix elements are arranged in order of increasing excitation energy, i.e., $\omega_1 \le \omega_2 \le \omega_3$, etc. Note that the $\chi_{\text{KS}}(\omega)$ matrix representation has two blocks, one representing the resonant part (excitation energies) and the other one the antiresonant part (de-excitation energies). 
In particular, the lowest KS excitation energy is given by $\omega_1 = \epsilon_{\text{CBM}} - \epsilon_{\text{VBM}}$, with $\epsilon_{\text{CBM}}$ and  $\epsilon_{\text{VBM}}$ the KS energy of the conduction band minimum (CBM) and the KS energy of the valence band maximum (VBM), respectively.

We now assume that also $\chi_0$ is diagonal in the Kohn-Sham basis.
This is an approximation, but it is in accordance with various practical calculations, in particular those based on the $GW$ method, in which $\chi_0$ is built with KS orbitals~\cite{AulburJonssonWilkins,Reining_book}. We can thus write
\begin{equation}
^4\chi_{0}(\omega) \!=\! 
\left(
\begin{array}{cccccc}
\!\frac{1}{\omega-\Omega_1} \!&  &  &  & & \\
&  \ddots  & & & & \\
&  & \!\frac{1}{\omega-\Omega_M}\! & & & \\
& &  & \! -\!\frac{1}{\omega+\Omega_1}\! & &  \\
&  & & & \ddots &\\
&  & & & &\!-\!\frac{1}{\omega+\Omega_M}\!
\end{array}
\right)
\end{equation}
%
%
where $\Omega_i$ are quasiparticle energy differences, 
i.e., differences of ionization potentials and electron affinities.
In particular, the lowest excitation energy is given by $\Omega_1 = I - A$, where $I$ 
is the first ionization potential and $A$ is the first electron affinity. Here we assume that the lowest excitation energy $\Omega_1$ 
corresponds to the same matrix element as the lowest KS excitation energy.
The four-point kernel ${^4}f^{(1)}_{xc} =  {^4}\chi^{-1}_{\text{KS}} - {^4}\chi_0^{-1}$ has hence the following simple frequency-independent matrix representation
%
\begin{equation}
{^4}f^{(1)}_{xc} \!=\!
\left(
\begin{array}{cccccc}
\!\Omega_1\!-\!\omega_1 \!\! &  &  &  & & \\
&  \ddots  & & & & \\
&  & \!\Omega_M\!-\!\omega_M \!\! & & & \\
& &  &\!\!\Omega_1\!-\!\omega_1  \!\!& &  \\
&  & & & \ddots &\\
&  & & & & \!\! \Omega_M\!-\!\omega_M\!
\end{array}
\right).
\label{Eqn:kernel1}
\end{equation}
%
%

The absorption onset is determined by the head of the matrix ${^4}f_{xc}^{(1)}$, which we will refer to in the following as ${^4}f_{xc,00}^{(1)}$.
It is given by
%
\begin{equation}
{^4}f_{xc,00}^{(1)}= I - A - (\epsilon_{\text{CBM}} - \epsilon_{\text{VBM}})
\label{Eqn:head}
\end{equation}
It can be shown that the ionization potential is exactly equal to minus the KS energy at the VBM, i.e., $I = -\epsilon_{\text{VBM}}$~\cite{Levy_1984,Almbladh_1985,Baerends_PCCP2017}.
Since an equivalent relation holds for the $N+1$ system, i.e., the system with
one additional electron, and the fact that $A$ should be equal to the first
ionization potential of the $N+1$ system, one can deduce that $A = -\epsilon_{\text{VBM}}^{N+1}$, where $\epsilon_{\text{VBM}}^{N+1}$ is the KS energy at the VBM of the $N+1$ system.
Therefore, we can rewrite Eq.~\eqref{Eqn:head} as
\begin{equation}
{^4}f_{xc,00}^{(1)} = \epsilon_{\text{VBM}}^{N+1} - \epsilon_{\text{CBM}}.
\label{Eqn:main1}
\end{equation}

Although Eq.~\eqref{Eqn:main1} seems a simple expression, it is not easy to calculate in practice. The problem arises from the fact that $\epsilon_{\text{VBM}}^{N+1}$ is difficult to evaluate in solids since they are usually described within the thermodynamic limit, which implies an infinite number of electrons from the outset.
However, $\epsilon_{\text{VBM}}^{N+1} - \epsilon_{\text{CBM}}$ 
is equal to $\Delta$, the difference between the fundamental gap $E_g$ 
and the KS gap $E_{\text{KS}}$ \cite{Perdew_1983,SS_83,Baerends_PCCP2017}.
This difference is also known as the derivative discontinuity~\cite{Perdew_1982,Perdew_1983}.
Therefore we arrive at the following relation,
\begin{equation}
{^4}f_{xc,00}^{(1)} = \Delta.
\label{Eqn:DELTA}
\end{equation}
This is one of the main results of this work.

Assuming a rigid shift of the conduction bands, $^4f_{xc}^{(1)}$ can thus be approximated 
by $^4f_{xc}^{(1)}=\Delta\, {^4}\textrm{I}$, where $^4\textrm{I}$ is the four-point identity matrix in transition space.
In practice, however, we prefer to use two-point quantities for numerical efficiency.
Therefore, it might be tempting to rotate ${^4}f_{xc}^{(1)}$ to a two-point kernel.
However, this would lead to a very complicated frequency-dependent quantity and would hence be very difficult to apply in practice \cite{Tokatly}. 
We will avoid this problem by including the effect of $^4f^{(1)}_{xc}$ on the four-point response function in transition space 
and only \emph{afterwards} rotating the latter to its two-point real-space representation.

We thus introduce a modified four-point Kohn-Sham polarizability ${^4}\chi^{(1)}_{KS}(\omega)$ defined by
%
\begin{equation}
[{^4}\chi^{(1)}_{\text{KS}}]^{-1}(\omega) = 
[{^4}{\chi}_{\text{KS}}]^{-1}(\omega) - \Delta\, {^4}\textrm{I} ,
\end{equation}
where the superscript $(1)$ indicates that $\chi^{(1)}_{KS}(\omega)$ contains $\Delta$.
%
Using the inverse of the basis set transformation in Eq.~\eqref{Eqn:basis_trans}, and taking the diagonal part of the resulting expression 
we obtain the following expression for the modified two-point KS polarizability,
%
\begin{equation}
\chi^{(1)}_{\text{KS}}(\mathbf{x}_1,\mathbf{x}_2,\omega) = \sum_{i,j}
\frac{(f_j - f_i) \phi_i(\mathbf{x}_1)\phi_j(\mathbf{x}_2)\phi^*_j(\mathbf{x}_1)\phi^*_i(\mathbf{x}_2)}{\omega - (\epsilon_i -\epsilon_j) -\text{sgn}(\epsilon_i -\epsilon_j)\Delta +i\eta}
\label{Eqn:modified-chi},
\end{equation}
which can be easily applied in practice.
The true response function can then be written in terms of $\chi^{(1)}_{\text{KS}}$ as
\begin{equation}
\chi(\omega)=\chi^{(1)}_{\text{KS}}(\omega)+\chi^{(1)}_{\text{KS}}(\omega)\left[
v_c+f_{xc}^{(2)}(\omega)
\right]
\chi(\omega),
\end{equation}
with $v_c$ the Coulomb potential.
From $\chi(\omega)$ one can readily obtain the inverse dielectric function $\varepsilon^{-1}(\omega) = 1 + v_c\chi(\omega)$.
The optical spectra are obtained from the imaginary part of $\epsilon^M(\omega)$, the macroscopic part of $\epsilon(\omega)$:
\begin{equation}
\varepsilon^M(\omega) = \varepsilon_1(\omega) + i \varepsilon_2(\omega).
\end{equation}

In order to apply our approach in practice we have to use an approximation for $\Delta$.
As first proposed by Kuisma \textit{et al.} \cite{Kuisma_PRB2010} and further discussed by Baerends \cite{Baerends_PCCP2017}, 
$\Delta$ can be approximated in terms of simple ground-state KS-DFT quantities according to
\begin{align}
\Delta &= K_{x}\sum_{i=1}^N
\left[\sqrt{\epsilon_{\text{CBM}}-\epsilon_i}-\sqrt{\epsilon_{\text{VBM}}-\epsilon_i}\right]
\notag \\ & \times 
\langle\phi_{\text{CBM}}| \frac{|\phi_i|^2}{\rho_0} |\phi_{\text{CBM}}\rangle
\label{Eqn:delta_GLLB}
\end{align}
where $\phi_{\text{CBM}}$ is the KS spinorbital corresponding to the CBM, $\rho_0$ is the ground-state density and $K_{x}=\frac{8\sqrt{2}}{3\pi^2}\approx 0.382$. 

The expression in Eq.~\eqref{Eqn:delta_GLLB} can be obtained from the GLLB (Gritsenko-van Leeuwen-van Lenthe-Baerends) approximation to the ground-state xc potential 
derived in Ref. [\onlinecite{GLLB_PRA1995}].
The GLLB functional is an approximation to the response part of the exact exchange optimized effective potential.
A detailed derivation of Eq.~\eqref{Eqn:delta_GLLB} is given by Baerends~\cite{Baerends_PCCP2017}.
The constant $K_{x}$ can be obtained from the uniform electron gas, where the GLLB exchange response potential becomes exact. \footnote{Note that one could consider variation of $K$ as a way to incorporate correlation effects (using
a $K_{xc}$).} Fundamental gaps calculated using the derivative discontinuity in Eq.~\eqref{Eqn:delta_GLLB} have been reported for a large number of solids. \cite{Kuisma_PRB2010,Baerends_PCCP2017,Thygesen_EESci2012,Thygesen_PRB2013,Thygesen_JPC2015}.
In general, the results are excellent. We note, however, that for gapped materials for which the KS band gap is zero, the correction in Eq.~\eqref{Eqn:delta_GLLB} is zero as well. One finds the same problem when calculating a $G_0W_0$ band gap on top of a KS metallic band structure.

In practice, we use a slight generalization of TDDFT, namely TD-current-DFT (TDCDFT)~\cite{DharaGhosh,GhoshDhara,Vignale,Sangalli_2017}.
The practical details of how we solve the KS equations within TDCDFT can be found elsewhere~\cite{Freddie,Arjan_2005,Pina,Arjan1,Arjan2,Pina2}. In particular our method works in real space with two-point quantities and we calculate the macroscopic polarization induced by a macroscopic electric field in terms of the macroscopic current. This allows us to have optical spectra with the correct optical gap and excitonic effects at the cost of an RPA calculation, i.e., $O(N^3)$ scaling instead of the $O(N^6)$ for methods working in transition space.
We approximate $f_{xc}^{(2)}$ with the polarization functional (PF) of Ref.~[\onlinecite{Berger_2015}] which accurately describes the excitonic effects in various systems.
We will refer to the full kernel, i.e.,  $f_{xc}^{(1,GLLB)} + f_{xc}^{(2,PF)}$, as the Pure kernel to highlight the fact that it is based on pure Kohn-Sham theory.
\begin{table} 
\caption{GLLB-SC values for $\Delta$ and fundamental gap E$_g$}
\begin{ruledtabular}
\begin{tabular}{lccc}
system & \, $\Delta$ [eV] & E$_g$ [eV] \\
\hline
Si & 0.38  &      1.12    \\
GaP&  0.76  &  2.48  \\
GaAs &  0.27 &   1.03 \\
C &   1.31 &    5.449 \\
LiF &   4.11 &    14.91 \\
Ar &   4.63 &    14.83 \\
\end{tabular}
\end{ruledtabular}
\label{Table1}
\end{table}
We implemented our approach in a modified version of the Amsterdam Density Functional (ADF) code ~\cite{ADF1,ADF2,ADF3}. We illustrate our approach by applying it to the calculation of the optical spectra of two very different classes of solids, standard semiconductors (Si, GaP, GaAs) and 
 wide-gap insulators (C, Ar, LiF).
We use the following lattice parameters: 5.43 \AA\,for Si, 5.42 \AA\,for GaP, 5.654 \AA\,for GaAs, 3.534 \AA\,for diamond, 4.026 \AA\,for LiF, and 5.26 \AA\,for solid argon. 
Moreover, we use the TZ2P (triple-$\zeta$ + 2 polarization functions) and QZ4P (quadruple-$\zeta$ + 4 polarization functions) basis sets provided by ADF for bulk silicon and LiF and for GaP, GaAs, diamond, and solid argon, respectively.
The $\mathbf{k}$-space integrals are done analytically using a Lehmann-Taut tetrahedron scheme~\cite{LehmanTaut}.
The ground-state calculations are done within the GLLB-SC xc potential ~\cite{Kuisma_PRB2010,GLLB_PRA1995,Baerends_PCCP2017}, 
which is based on the PBEsol \cite{PBEsol} correlation potential and uses the GLLB approximation to the exchange optimized effective potential. For GaP and GaAs we also include scalar relativistic effects.
The GLLB-SC values we obtained for $\Delta$ and the fundamental gap are reported in Table \ref{Table1}. 
We note that the approximation $f_{xc}^{(2,PF)}$ induces a gap renormalization \cite{Sottile_PRL2003,Marini_PRL2003}; this is clear from the example of solid argon for which we have an optical gap of about 11.6 eV, which is about 3 eV smaller than the (direct) fundamental gap.

%

\begin{figure}[t]
\centering
\includegraphics[width=0.8\columnwidth]{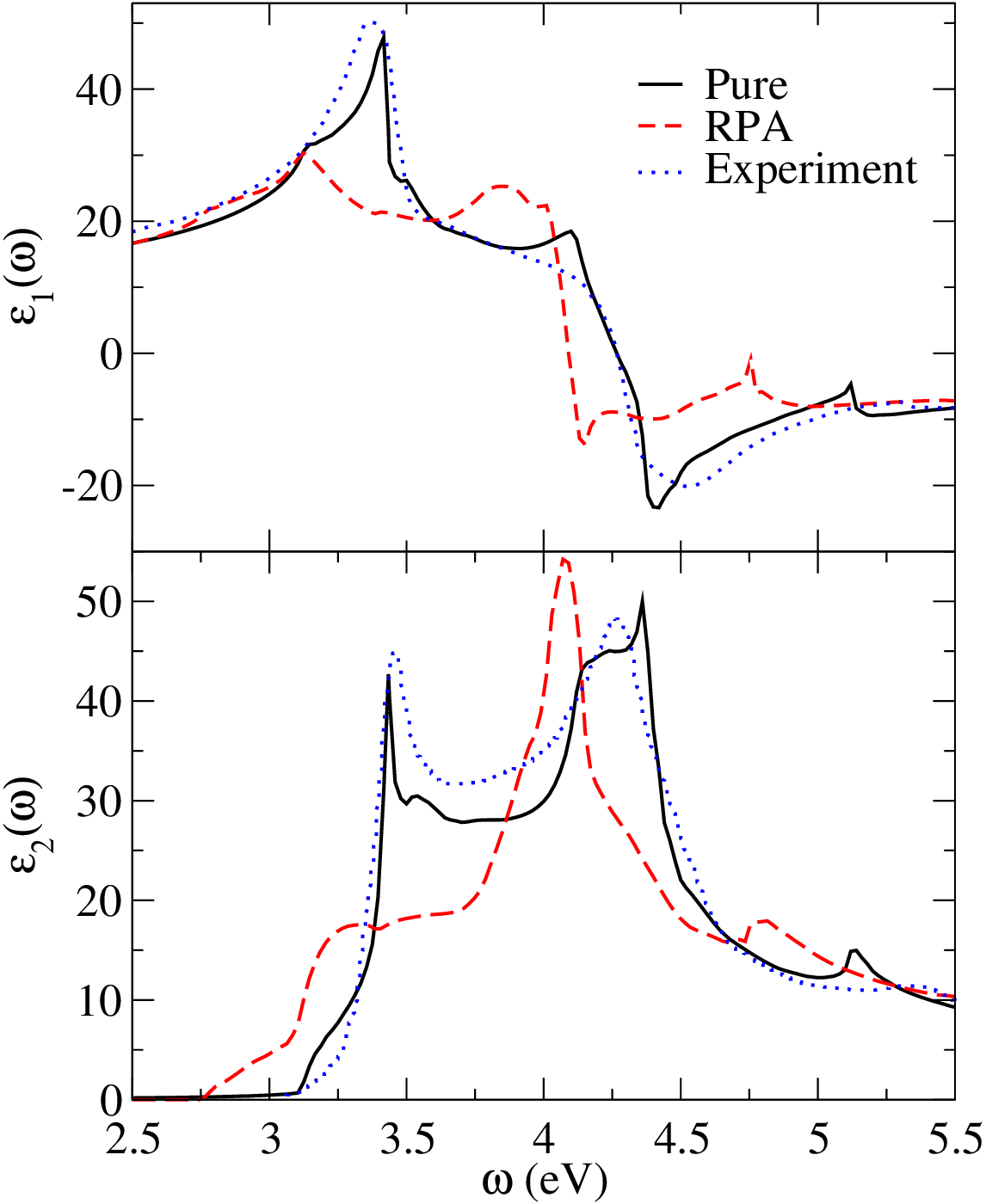}
\caption{(Color online) Real ($\epsilon_1(\omega)$) and imaginary ($\epsilon_2(\omega)$) parts of the dielectric function of bulk silicon.
Solid line (black): Pure functional (Pure); Dashed line (red): RPA; 
Dotted line (blue): experiment from Ref.\ [\onlinecite{Lautenschlager_sil}].}
\label{Fig:Si}
\end{figure}

\begin{figure}[t]
\centering
\includegraphics[width=0.8\columnwidth]{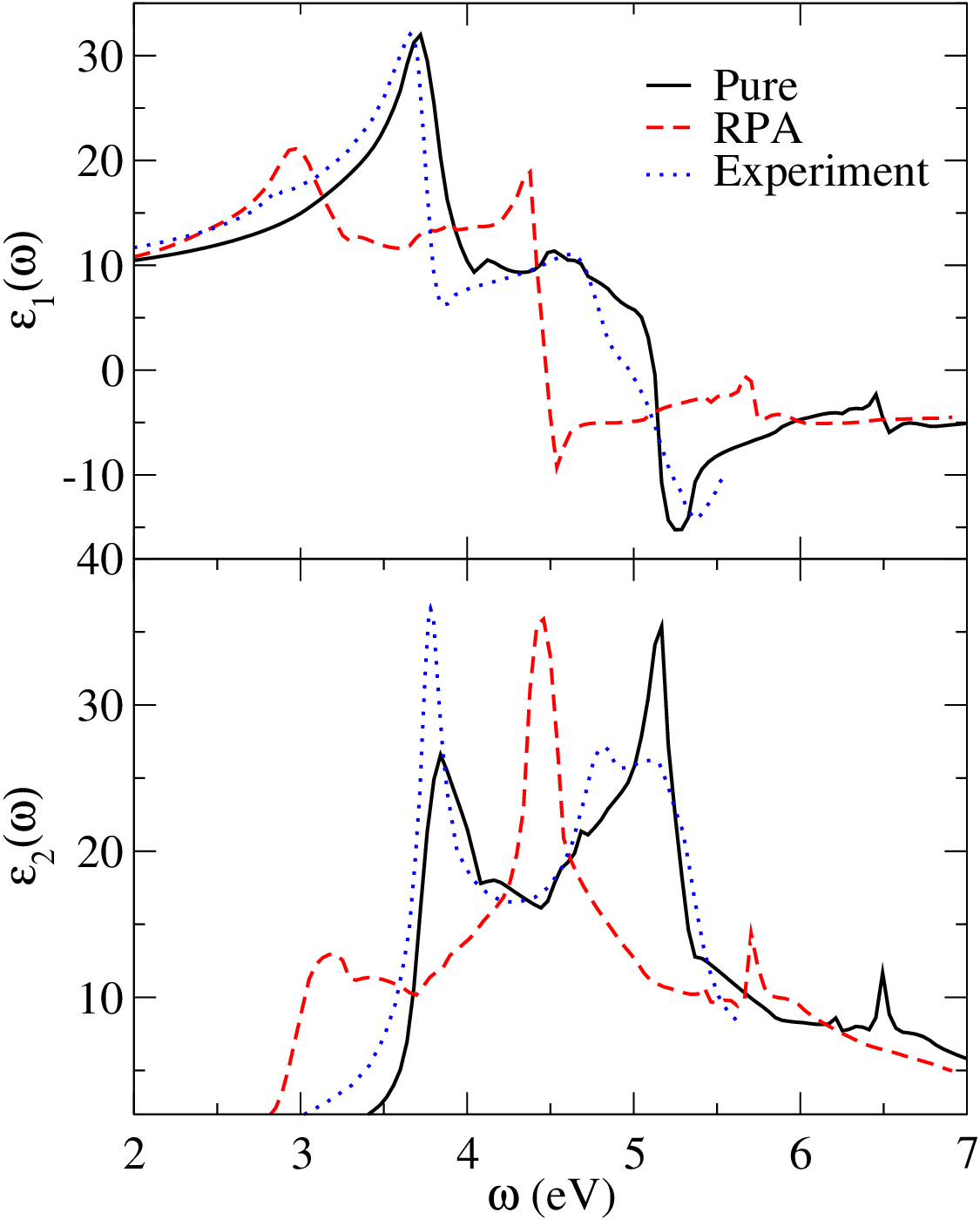}
\caption{(Color online) Real ($\epsilon_1(\omega)$) and imaginary ($\epsilon_2(\omega)$) parts of the dielectric function of bulk GaP.
Solid line (black): Pure functional (Pure); Dashed line (red): RPA; 
Dotted line (blue): experiment from Ref.\ [\onlinecite{EXP_GaP}].}
\label{Fig:GaP}
\end{figure}

\begin{figure}[t]
\centering
\includegraphics[width=0.8\columnwidth]{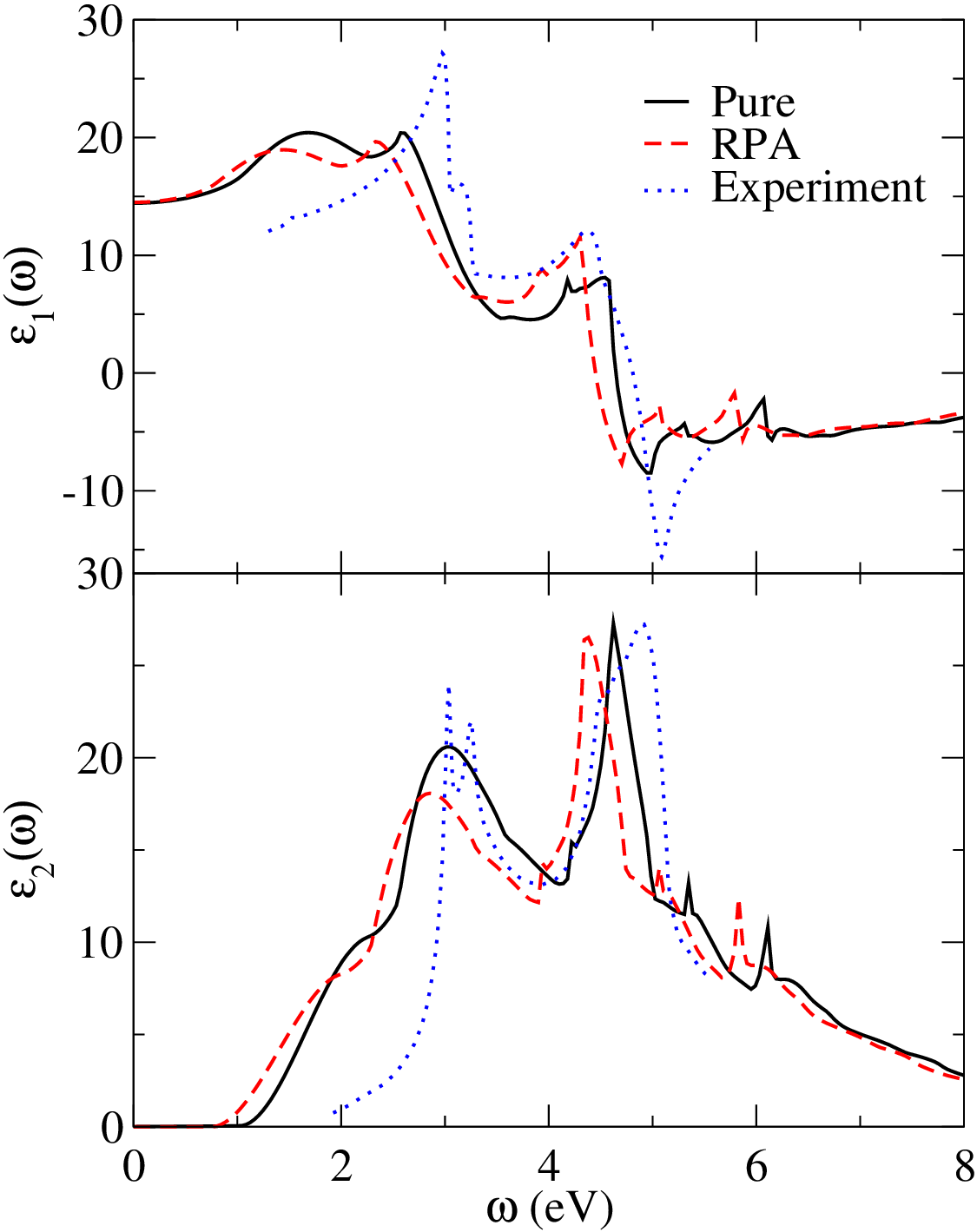}
\caption{(Color online) Real ($\epsilon_1(\omega)$) and imaginary ($\epsilon_2(\omega)$) parts of the dielectric function of bulk GaAs.
Solid line (black): Pure functional (Pure); Dashed line (red): RPA; 
Dotted line (blue): experiment from Ref.\ [\onlinecite{EXP_GaAs}].}
\label{Fig:GaAs}
\end{figure}

In Figs.~{\ref{Fig:Si}}, {\ref{Fig:GaP}}, and {\ref{Fig:GaAs}} we report the dielectric function of bulk Si, GaP, and GaAs, respectively,
calculated with the Pure functional at 0 Kelvin and compare it to the RPA spectrum as well as to the experimental spectrum obtained at 30 Kelvin for Si, 15 Kelvin for GaP, and 22 Kelvin for GaAs. 

\begin{figure}[t]
\centering
\includegraphics[width=0.8\columnwidth]{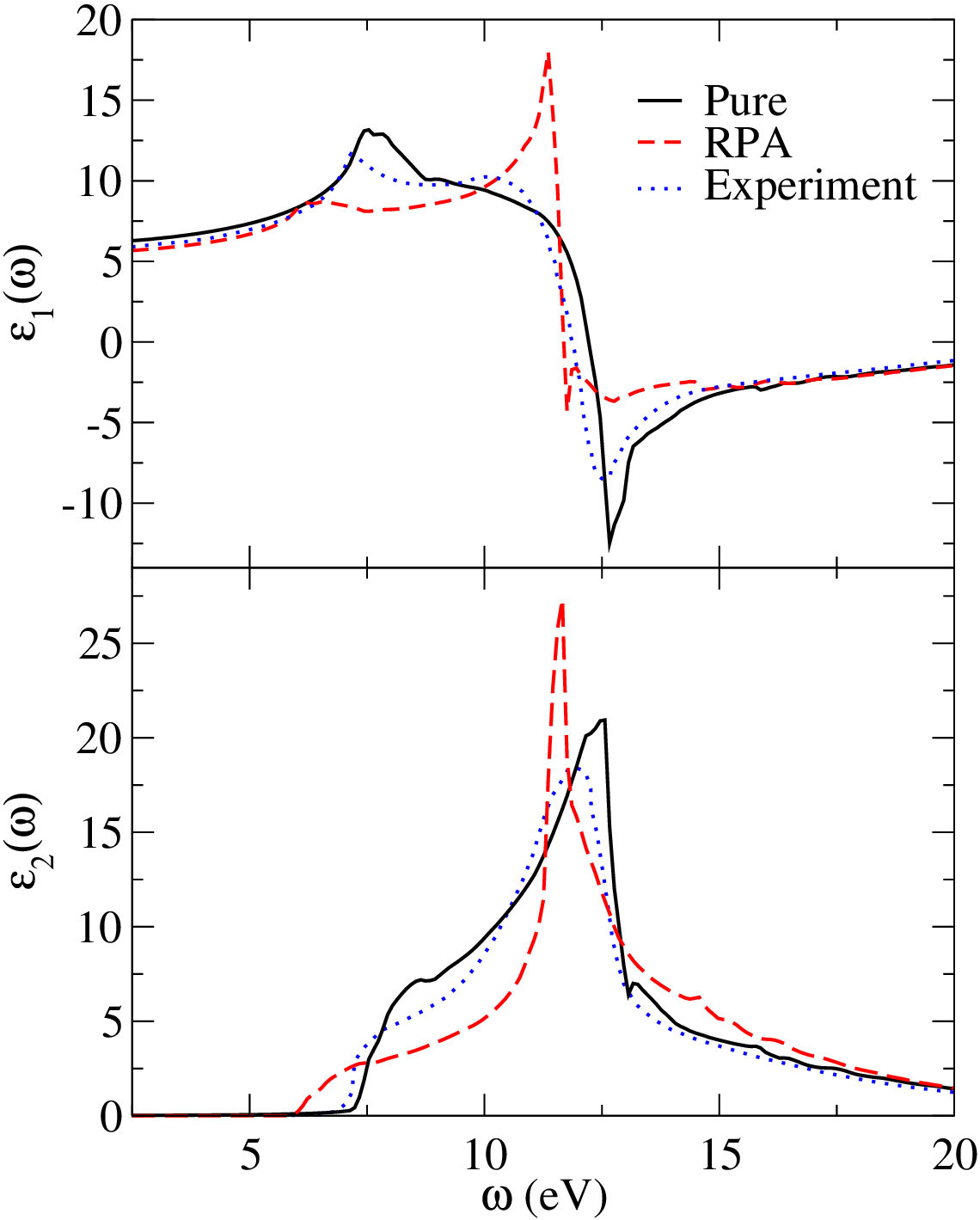}
\caption{(Color online) Real ($\epsilon_1(\omega)$) and imaginary ($\epsilon_2(\omega)$) parts of the dielectric function of diamond.
Solid line (black): Pure functional (Pure); Dashed line (red): RPA; 
Dotted line (blue): experiment from Ref.\ [\onlinecite{EXP_C}].}
\label{Fig:C}
\end{figure}

\begin{figure}[t]
\centering
\includegraphics[width=0.8\columnwidth]{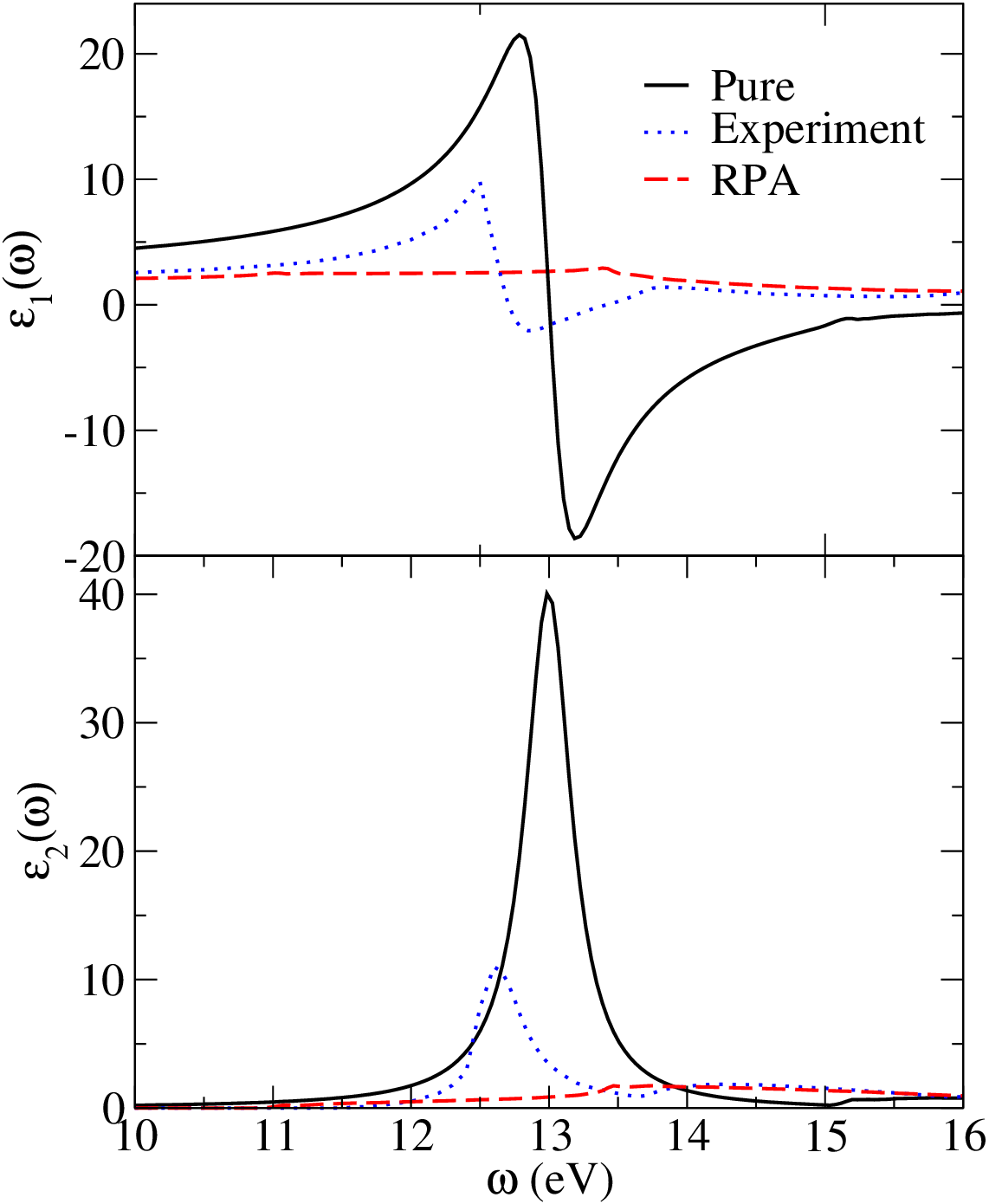}
\caption{(Color online) Real ($\epsilon_1(\omega)$) and imaginary ($\epsilon_2(\omega)$) parts of the dielectric function of bulk LiF.
Solid line (black): Pure functional (Pure); Dashed line (red): RPA; 
Dotted line (blue): experiment from Ref.\ [\onlinecite{Roessler:67}].}
\label{Fig:LiF}
\end{figure}

\begin{figure}[t]
\centering
\includegraphics[width=0.8\columnwidth]{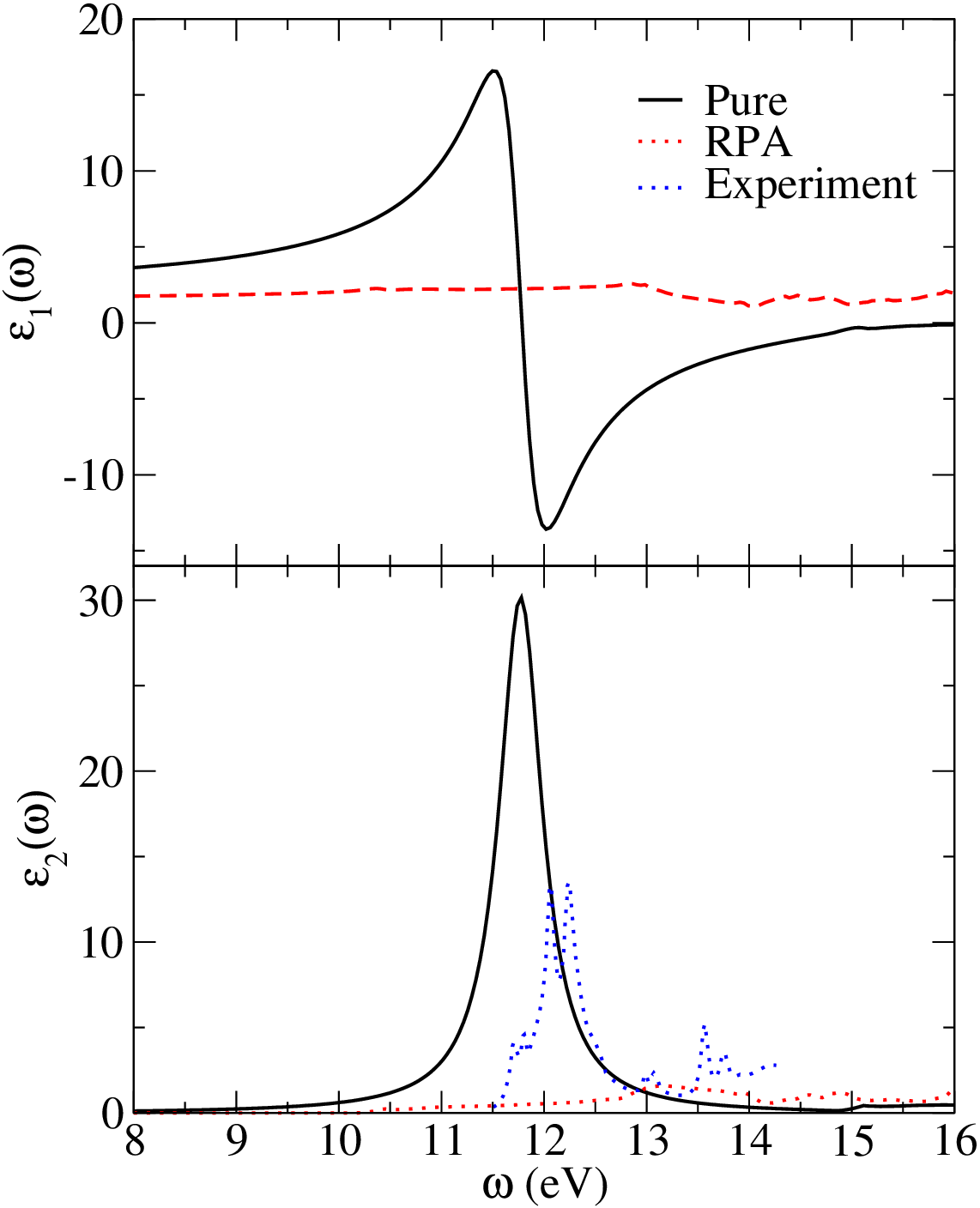}
\caption{Color online) Real ($\epsilon_1(\omega)$) and imaginary ($\epsilon_2(\omega)$) parts of the dielectric function of solid argon.
Solid line (black): pure functional (Pure); Dashed line (red): RPA; 
Dotted line (blue): experiment from Ref.\ [\onlinecite{Saile_Ar}].}
\label{Fig:Ar}
\end{figure}

The Pure kernel yields an absorption spectrum that is in good agreement with the experimental measurements for the three systems. In particular, the peak positions and the excitonic effects are well reproduced. Also, the real part of the dielectric function obtained with the Pure kernel compares well to the experiment.
Instead, the RPA spectrum exhibits the well-known shortcomings mentioned before, i.e., the underestimation of the absorption onset and the absence of excitonic effects. In case of GaAs the RPA spectra are already in overall good agreement with experiment, and the Pure kernel only slightly improves the exciton position and intensity. Note that the splitting of the first peak in the experimental spectrum of GaAs is not reproduced in the calculated spectrum, because it is due to spin-orbit coupling, which is neglected in our calculations. 
We note that the theoretical spectra have more structure than the experimental spectrum because it is calculated at 0 Kelvin and no broadening parameter is used to simulate temperature effects. 
In Figs.~{\ref{Fig:C}}, {\ref{Fig:LiF}}, and {\ref{Fig:Ar}} we show the dielectric function of diamond, LiF, and solid argon, respectively, calculated with the Pure kernel at 0 Kelvin and compare it to the RPA spectrum as well as to the experimental spectrum obtained at room temperature. We see that the onset of the absorption is well-reproduced as is the full spectrum, except for an overestimation of the intensity of the bound exciton in LiF and solid Ar. We note that this overestimation is common to similar kernels derived to describe excitonic effetcs \cite{santiago,bootstrap}. Unfortunately, to the best of our knowledge, there is no experimental data of $\epsilon_1(\omega)$ of solid Ar.

In conclusion, we have made explicit the link between the derivative discontinuity of
ground-state DFT and the xc kernel of TDDFT.
Using this link we proposed the Pure kernel, which combines the derivative discontinuity and the polarization functional, to describe optical spectra.
We showed that it yields optical spectra in good agreement with experiment for typical examples of standard semiconductors and wide-gap insulators.  The central issue here is that these results were obtained within a pure KS approach without resorting to empirical data or approaches that go beyond TDDFT. Finally we note that the kernel we propose in this work is an approximation to the TDDFT kernel $f_{xc}$, it hence can in principle be used to calculate all the properties TDDFT can access. In particular, it could be useful in total energy calculations.

This work has been supported through the EUR grant NanoX ANR-17-EURE-0009 in the framework of the ``Programme des Investissements d'Avenir'' and by ANR (project no. ANR-18-CE30-0025-01).
%
%

\end{document}